\documentclass[a4paper,12pt]{article}
\usepackage{graphicx,epsf}

\begin{document} 
\title{From heaviness to lightness during inflation 
} 
 
\vskip 1cm 
\author{David Langlois$^1$ and  Filippo Vernizzi$^{1,2}$\\ 
{\small {\it 1) GReCO, Institut d'Astrophysique de Paris, CNRS,}}\\ 
{\small {\it 98bis Boulevard Arago, 75014 Paris, France}}\\ 
{\small {\it 2) Helsinki Institute of Physics,
P.O. Box 64 (Gustaf Hallstromin katu 2),}}\\ 
{\small {\it FIN-00014 University of Helsinki - Finland}}}

\maketitle 
 
\def\beq{\begin{equation}} 
\def\eeq{\end{equation}} 
\newcommand{\bea}{\begin{eqnarray}} 
\newcommand{\eea}{\end{eqnarray}} 
\def\bi{\begin{itemize}} 
\def\ei{\end{itemize}} 
\def\V{U} 
\def\ti{t_{\rm in}} 
\def\taui{\tau_{\rm in}} 
\def\ai{a_{\rm in}} 
\def\x{\bf x} 
\def\m{m} 
\def\mo{m_\phi} 
\def\qi{q_{\rm in}}
\def\Phii{\Phi_{\rm in}}
\def\q{{q}}
\newcommand{\gsim}{\ \raise.3ex\hbox{$>$\kern-.75em\lower1ex\hbox{$\sim$}} \ } 
\newcommand{\lsim}{\ \raise.3ex\hbox{$<$\kern-.75em\lower1ex\hbox{$\sim$}} \ } 
 
\abstract{We study the  
quantum fluctuations of scalar fields with  
a variable effective mass during an inflationary phase. We consider the  
situation where the effective mass depends on a background scalar 
field, which  evolves  
 during inflation from being frozen into a damped  oscillatory phase when  
the Hubble parameter decreases below its mass. 
We find  power spectra with 
suppressed amplitude on  large scales, similar 
to the standard massless spectrum on small scales, and affected by 
modulations on intermediate scales. 
We stress the analogies and differences 
with the parametric resonance in the preheating scenario. We also discuss
some potentially observable consequences when the scalar field behaves
like a curvaton.}

\section{Introduction} 
The quantum fluctuations of scalar fields during inflation are one  
of the cornerstones of modern early cosmology (see e.g. \cite{textbooks,MFB}  
for textbook reviews). In the standard inflationary 
paradigm, the quantum fluctuations of the inflaton field are directly  
responsible for the cosmological perturbations, which today can be observed  
 as CMB fluctuations.  
In some recent variants of the inflationary scenario,  the inflaton  
fluctuations are not related  to the cosmological perturbations, this r\^ole  
being  taken over by the {\it curvaton}, a light scalar field whose quantum  
fluctuations generated {\it during inflation}, ultimately   
give birth to the primordial cosmological perturbations \cite{lw02,es02,mt01}. 
 
In both cases mentioned above, the crucial ingredient for generating  
cosmologically  relevant perturbations is {\it lightness}: the effective mass 
of the scalar field must be much smaller than the Hubble parameter during  
inflation. By contrast, the  fluctuations generated during inflation 
by very massive scalar fields,  
with $m\gg H$,  
 are suppressed. 
 
The purpose of the present work is to explore the quantum fluctuations  
of scalar fields whose effective mass is varying with time during inflation. 
The effective mass is due to the coupling to  a scalar field with a non zero 
background value  and  
can thus change significantly  
during inflation as a consequence  of the cosmological evolution of this  
background scalar field, which is first frozen at some non zero value and then 
oscillates around the minimum of its potential.

The effect  of an oscillating mass on quantum fluctuations has   
been thoroughly studied in the context of the so-called ``preheating'' 
scenario, which is supposed to take place just after the end of inflation 
when  
the inflaton oscillates at the bottom of its potential \cite{KLS}.  
This is known to lead to an amplification of the quantum fluctuations 
via parametric resonance. 
In the present work, however, we consider the amplification of quantum 
fluctuations  {\it during inflation} and not after inflation as in the  
preheating scenario.  
The effect of an oscillating mass during inflation has also been  
discussed  in \cite{Burgess:2002ub} in the context of hybrid inflation.  
In that work,  it is the inflaton field which  
has an oscillating mass due to a coupling 
to the field that triggers the end of inflation.  
In our case, by contrast, 
the field with a time-dependent mass is not the inflaton but a spectator 
field during inflation.  
Thus, in both cases, even if the basic equations are equivalent, the  
regimes under investigation are rather different.  

This paper is organized as follows.  In the next section,  
we discuss quantum fluctuations 
of a massive scalar field during de Sitter inflation.  
In Sec.~\ref{sec:3}, we present the model and derive the equations for the  
evolution of the quantum fluctuations of a field whose mass  
depends on the expectation value of an oscillating scalar field. We also 
briefly review the preheating scenario. In  Sec.~\ref{sec:4}, we solve 
numerically the equation governing the field fluctuations and 
 discuss the corresponding power spectrum.
In Sec.~\ref{sec:5}, as  
an example of an inflationary phase with a decreasing Hubble parameter,  
we consider power-law inflation. Finally, some 
possible  
observational consequences  are discussed in Sec.~\ref{sec:6}.  
 
\section{Quantum fluctuations of massive\\ scalar fields} \label{sec:2} 
We start by recalling the usual derivation of quantum fluctuations for  
a test  
scalar field, $\chi$, with a {\it constant mass} $m_\chi$, during inflation. 
Its action is given by 
\beq 
S=\int d^4x \sqrt{-g}\left[ -{\textstyle {1 \over 2}}\chi_{,\mu} 
\chi^{,\mu} - {\textstyle {1 \over 2}} m_\chi^2 \chi^2\right]. 
\eeq 
The homogeneous component is governed by the Klein-Gordon equation in  
 a Friedmann-Lema\^\i{}tre-Robertson-Walker (FLRW) spacetime with metric 
\beq 
ds^2=-dt^2+a(t)^2 d{\x}^2, 
\eeq 
which is given by  
\beq 
\ddot\chi+ 3H \dot\chi+ m_\chi^2\chi=0, 
\eeq 
where $H\equiv \dot a/a$ is the Hubble parameter. 
In a quantum framework, it is useful to Fourier decompose the  
fluctuations and to write  
\begin{equation} 
\hat {\delta\chi}(t, {\bf x}) = 
{1\over{(2\pi)^{3/2}}} \int d^3k\ {\Bigl( \hat a_{{\bf k}} \delta\chi_{ 
k}(t)\, e^{ - 
i{{\bf k}}{{\bf x}}} 
+ \hat a_{{\bf k}}^ + \delta\chi^*_k(t)\, e^{i{{\bf k}}{{\bf x}}} 
\Bigr)}\ , 
\label{37} 
\end{equation} 
where $\hat a_{{\bf k}}$ and $\hat a_{{\bf k}}^ + $  
are respectively annihilation and creation operators. 
The equation of motion for $\delta\chi_k$ is given by 
\beq 
\ddot {\delta\chi}_k + 3{{\dot a}\over a}\dot{\delta\chi}_k + {\left( 
{{ k^2}\over a^2} 
 + m^2_{\chi}  \right)} \delta\chi_k = 0 \ . 
\label{KG} 
\end{equation} 
 
For simplicity, in this section we consider only the
case of de Sitter inflation.  
During a de Sitter phase, the Hubble parameter $H$ is strictly constant  
and 
\beq 
a(t)=e^{Ht}. 
\eeq 
Introducing, instead of the cosmic time $t$, the conformal 
time defined by $d\tau=dt/a(t)$, so that  
\beq 
a(\tau)=-{1\over H\tau}, 
\eeq 
and  the new function
\beq 
u_k \equiv a \delta\chi_k, 
\eeq 
the  equation of motion  (\ref{KG}) is replaced by   
\beq 
\label{eq_u_dS} 
u_k''+\left[k^2-{2-(m_\chi^2/H^2)\over \tau^2}\right]u_k=0.  
\eeq 
We have implicitly assumed a spatially flat gauge, where
the variable $u$ defined here coincides with the Sasaki-Mukhanov 
variable \cite{MFB} and, since $\chi$ is a test field, we have neglected 
 the
self-gravity, i.e., the backreaction of metric perturbations on the 
field fluctuations, in Eq.~(\ref{KG}).
The general solution of Eq.~(\ref{eq_u_dS}) is given by 
\beq 
u_k(\tau)=\sqrt{-k\tau}\,  B_\nu(-k\tau),  
\qquad \nu^2=\frac{9}{4}-\frac{m_\chi^2}{H^2}, 
\eeq 
where $B_\nu$ is a linear combination of Bessel functions of order $\nu$ 
\cite{AS}. 
The appropriate linear combination is determined by the  
requirement that, for wavelengths much  smaller than the Hubble radius,  
i.e. $k|\tau|\gg 1$, the quantum state must correspond to that of the usual  
Minkowski vacuum, i.e. $u_k \simeq e^{-ik\tau}/\sqrt{2k}$.  
Using the asymptotic behavior of Bessel functions,  
in particular 
\beq 
H_\nu^{(1)}(z)\sim \sqrt{2\over \pi z}e^{i\left(z-{\pi\over 2}\nu 
-{\pi\over 4}\right)}, 
\quad  
H_\nu^{(2)}(z)\sim \sqrt{2\over \pi z}e^{-i\left(z-{\pi\over 2}\nu 
-{\pi\over 4}\right)}, 
\quad  
|z|\gg 1, 
\eeq 
this  
implies that the appropriate linear combination is given by  
\beq 
\label{vacuum} 
u_k(\tau)= 
{\sqrt{\pi}\over 2}e^{i\left({\pi\over 2}\nu+{\pi\over 4}\right)} 
\sqrt{-\tau}\,  
H_\nu^{(1)}(-k\tau). 
\eeq 
Within the Hubble radius ($k|\tau|\gg 1$), i.e. at early-times/small-scales, 
the solution $u_k$ is thus oscillating with constant amplitude. 
In the opposite limit $k|\tau|\ll 1$, on super-Hubble scales,  
 the asymptotic behavior is given by 
\beq 
H_\nu^{(1)}(z)\sim -{2^\nu i\Gamma(\nu)\over\pi}z^{-\nu}, 
\quad |z|\ll 1. 
\eeq 
 
If we define the power spectrum of $\delta\chi$ as 
\beq 
{\cal P}_{\delta\chi}(k) \equiv {k^3\over 2\pi^2}|\delta\chi_k|^2
={k^3\over 2\pi^2}{|u_k|^2\over a^2} 
, \label{eq:power_spectrum} 
\eeq 
we find 
\beq 
{\cal P}_{\delta\chi}(k)= {H^2\over 8\pi} \left(k|\tau|\right)^3\left|H_\nu^{(1)} 
(k|\tau|)\right|^2. 
\eeq 
On super-Hubble scales, this spectrum behaves like 
\beq 
{\cal P}_{\delta\chi}(k)\simeq {2^{(2\nu-3)}\Gamma^2(\nu)\over \pi^3}H^2 
 \left(k|\tau|\right)^{3-2\nu}. 
\eeq 
 
For  
a {\it massless} scalar field, the most important case considered  
in the literature, $m_\chi=0$  
and   $\nu = 3/2$ and one recovers, on large scales, 
 the famous scale-invariant spectrum, 
\beq 
{\cal P}_{\delta\chi}(k) 
\simeq  \left({H\over 2\pi}\right)^2  
\qquad (k\ll aH).  
\eeq 
A {\it light} scalar field, $m_\chi\ll H$, behaves similarly to the massless 
one. 
By contrast, for a {\it heavy} scalar field with  
$m_\chi \gg H$, the spectrum is suppressed in the limit $k|\tau|\ll 1$.

\section{Coupled scalar fields} \label{sec:3} 
\subsection{The model} 
After having considered a single scalar field with a constant mass, let us  
now study the case of a test scalar field $\chi$ whose mass depends on  
the expectation value of another scalar field, say $\phi$. To be more  
explicit, let us consider the specific model defined by the Lagrangian 
\beq 
\label{lagrangian} 
L = -{\textstyle {1 \over 2}} \phi_{,\mu} \phi^{,\mu} - {\textstyle {1\over 2}} 
\mo^2\phi^2 - 
{\textstyle {1 \over 2}} 
\chi_{,\mu} 
\chi^{,\mu} - {\textstyle {1 \over 2}} m_\chi^2 \chi^2  
 - {\textstyle {1 \over 2}} g^2 \phi^2 \chi^2. 
\eeq 

In the early universe,  these
two scalar fields can evolve according to the following scenario. 
We first assume that there is 
 a phase of inflation, during which these two scalar fields  
are simply spectator fields, i.e. their total energy density is negligible  
with respect to that of the field responsible for inflation. Taking some  
arbitrary initial   values  
for the two scalar fields $\phi$ and $\chi$, let us discuss  
their subsequent evolution.  
In a first phase, assuming that the  
Hubble parameter $H$ is bigger than the mass $\mo$, the scalar field  
$\phi$ is frozen at its initial value $\phi_0$.  
This implies that the effective  
(squared) mass of $\chi$ is $m_{\rm eff}^2=m^2_{\chi}  + g^2\phi_0^2$,  
potentially much bigger than its bare mass $m_\chi$ (which will  
be taken to be zero in practice in this work). If   
$m_{\rm eff}\gg H$, then the   
scalar field  
expectation value  
$\chi(t)$ quickly rolls down to zero.  
 
Subsequently, the  
Hubble parameter in the universe decreases, either slowly like  
in slow-roll inflation, or abruptly if there are phase transitions.  
If, {\it during inflation}, the Hubble parameter decreases {\it below} 
the mass $\mo$  then the scalar field  
$\phi$ 
will start moving from its initial value and will then 
 oscillate about the minimum 
of its potential (here $\phi=0$). This will induce an {\it oscillating} 
effective mass for the scalar field $\chi$, in particular for its  
fluctuations, since the background value is now zero. Our purpose is  
to compute the spectrum of the fluctuations generated by this time-dependent 
mass.  
 
In a FLRW spacetime, the equation of motion for the Fourier modes  
$\delta\chi_k$ 
becomes 
\beq 
\ddot{\delta\chi}_k + 3{{\dot a}\over a}\dot{\delta\chi}_k + {\left( 
{{ k^2}\over a^2} 
 + m^2_{\chi}  + g^2\phi^2 \right)} \delta\chi_k = 0 \ ,  
\label{eq:fluctuation_phi} 
\eeq 
which means that the effective, time-dependent,  square mass is given by  
\beq 
m_{\rm eff}^2(t)= m^2_{\chi}  + g^2\phi^2(t). 
\eeq 
Note that this is exactly the equation of motion that has been studied  
in the context of the preheating scenario \cite{KLS}.  
In contrast to the preheating  
scenario where the scalar field $\phi$ is the inflaton which, after inflation, 
oscillates at the bottom of its potential, we  consider the above  
equation {\it during inflation}.  This means that $\phi$ cannot be the 
inflaton 
field, since the effective equation of state of an oscillating massive  
scalar field is that of pressureless matter,  
 and thus inflation must be supported by another scalar field, which  
we do not need to specify  here, or possibly by a different mechanism. 
We will focus first on the simplest case where the background is characterized 
by an exact de Sitter phase, i.e. for which $H$ is strictly constant. Later 
we will consider the more realistic case where the Hubble parameter decreases slowly. 
 
The equation of motion for the scalar field $\phi$ is given by  
\beq 
\label{KG_phi} 
\ddot\phi+ 3H \dot\phi+ \mo^2\phi=0. 
\eeq 
 For $\mo>3H/2$, the general solution  is given by 
\beq 
\phi=\lambda \exp\left(-{3\over 2}Ht\right) 
\cos\left(\m t-\varphi\right), 
\quad \m \equiv \mo\sqrt{1-{9H^2\over 4\mo^2}}. 
\eeq 
We choose an initial time $\ti$ such that $\dot \phi(\ti)=0$. The  
solution can then be written in the form  
\beq 
\label{sol_phi} 
\phi=\Phii \exp\left(-{3\over 2}H(t-\ti)\right) 
\cos\left[\m(t-\ti)-\varphi\right], \quad  
\varphi\equiv \sin^{-1}\left({3H\over 2\mo}\right). 
\eeq 
Note that the above definition of the phase $\varphi$ implies  
$\m=\mo \cos\varphi$. 
To describe the transition from a frozen field  
$\phi=\Phii \cos \varphi$,  
during an  
early inflationary phase with $H>\mo$, into an oscillatory behavior, one  
would need a decreasing Hubble parameter, as mentioned above. This will be  
considered in Sec.~\ref{sec:5} by assuming power-law inflation.

As recalled earlier, when the mass of the scalar field is bigger   
than the Hubble parameter, the fluctuation spectrum is suppressed. 
The case we wish to investigate now is more complicated because, due 
to the coupling, $\chi$ has an effective mass that is changing  
with time, 
\beq 
m_{\rm eff}^2=m_{\chi}^2 
+ g^2\Phii^2e^{-{3H\Delta t}}\cos^2(\m\Delta t-\varphi), 
\eeq 
where $\Delta t=t-\ti$. 
 
 As in the previous section, it is possible
 to rewrite the equation of motion  
for the fluctuations, Eq.~(\ref{eq:fluctuation_phi}),  
in terms of the function $u_k=a \delta\chi_k$ while using 
the conformal time $\tau$.  
The equation of motion now reads 
\beq 
{d^2 u_k\over d\tau^2}+\left(k^2- \V(\tau)\right) u_k=0, 
\quad \V(\tau)= {a''\over a}-m_\chi^2 a^2 -g^2\phi^2 a^2. 
\label{eq_u} 
\eeq 
In the case of a de Sitter background, the scale factor is given  
by  
\beq 
a=e^{H(t-\ti)}=-{1\over H\tau}, 
\eeq 
where our choice of normalization is such that $a(\ti)=1$, which implies  
in particular that $\taui=-1/H$. 
The effective potential in the perturbation equation is then given by  
\beq 
\label{potential} 
\V(\tau)= {2\over \tau^2} -  {m_\chi^2\over H^2\tau^2}  
+g^2\Phii^2 H\tau\,  \cos^2\left[{{\m}\over H}\ln(-H\tau)+\varphi\right]. 
\eeq 
Equation~(\ref{eq_u}) represents  
the equation of an harmonic oscillator with a time-dependent frequency 
$\omega_k(\tau)$, defined  as 
\beq  
\omega_k^2(\tau)\equiv k^2 - \V(\tau) 
. 
\eeq 
Numerical integration of the above equation will be performed in the next  
section. Before, it is instructive  to review briefly the  
preheating scenario, which will turn out to  be useful to interpret our  
results.

\subsection{Preheating} 
Parametric resonance during preheating has been studied in detail  
in \cite{KLS}. There, $\phi$ is the inflaton field, which oscillates  
 after the end of  
inflation and $\chi$ is a test scalar field  coupled to the inflaton as in  
our Lagrangian  
(\ref{lagrangian}). As a consequence,  
the equations of motion for preheating are 
the same as in our scenario, equations (\ref{KG_phi}) and  (\ref{eq_u}).

Following \cite{KLS}, it is easier to  discuss first  
the equations when expansion is ignored, for  $a=1$ (and $\m = \mo$).   
The scalar field $\phi$ then oscillates without damping (at least as long 
as backreaction can be ignored) and  Eq.~(\ref{eq_u}) reduces to  
an equation of the form 
\beq 
{d^2 u_k\over d\tau^2}+\left(k^2+ g^2 \Phi^2 \cos^2(\m \tau)  \right) u_k=0, 
\eeq 
where, for simplicity, we  have assumed  $m_\chi=0$. 
 
This equation can be rewritten as a Mathieu equation describing an oscillator 
with a periodically changing frequency, 
\begin{equation} 
\frac{d^2 u_k}{d z^2} + \left(A_k + 2\q \cos 2z \right) u_k = 0 \ , 
\label{eq:Mathieu} 
\end{equation} 
with 
 \beq z= \m \tau , \quad  
A_k 
= {k^2 \over \m^2 }+2\q, 
\eeq 
where $\q$ is defined by 
\beq 
\q \equiv \frac{g^2 \Phi^2}{4 \m^2}. \label{eq:q_def} 
\eeq

In the {\it broad resonance regime}, characterized  
by $\q \gg 1$, one observes an abundant production of particles  
when the amplitude of $\phi$ goes through zero. This essentially  
 corresponds to the  
times when adiabaticity is broken, i.e. when 
\beq 
\left| {d\omega_k\over d\tau}\right|\gsim  \omega_k^2. 
\eeq 
Substituting $\omega_k^2=k^2+g^2\phi^2$ in the above  condition, 
one finds that  
a non-adiabatic regime occurs for the modes satisfying  \cite{KLS}
\beq 
k \lsim \m \q^{1/4}. 
\eeq 
 
When expansion is taken into account, the scalar field amplitude decreases  
with time. This implies that the effective parameter $q$, now 
time-dependent, rapidly decreases  
during the cosmological evolution, which progressively makes the broad  
resonance more and more narrow.  In the context of the preheating  
scenario this leads to {\it stochastic resonance}.

\section{Spectrum of perturbations} \label{sec:4} 
 
In the present section, we discuss the numerical  
resolution of the second order differential equation (\ref{eq_u}) 
with the potential (\ref{potential}) whose shape is plotted on Fig. 1. 
For simplicity, we will consider only the case $m_\chi=0$. 

\subsection{Initial conditions} 
We must first specify the initial conditions that we use in our numerical  
integration. As discussed in the previous section, our initial (conformal) 
time is $\taui= -1/H$, so that $\ai=1$. 
The  initial conditions for  
$\phi$ and its first time derivative are 
\beq  
\phi 
=\Phii\cos\varphi, \quad \dot \phi = 0,  
\quad \rm{ at } \ \ \tau=\taui. \label{eq:ini1} 
\eeq 
 
For the perturbations $u_k$, we would like to take an  initial state  
that corresponds to a  {\it vacuum}. However, a vacuum is unambiguously 
defined only in some asymptotic regimes. In the  
present case, the effective potential $\V(\tau)$ oscillates 
with increasing amplitude as one goes backwards in time. However, this  
behavior is an artifact of our simplified model since in the more realistic  
situation, mentioned  earlier,  
 the scalar field $\phi$ is initially frozen, so that the effective mass  
$m_{\rm eff}$ is {\it constant} in the asymptotic past. Consequently, one 
can choose  initial conditions  defined by taking at time $\tau= \taui$ the  
solution (\ref{vacuum}), corresponding to the vacuum in the  
asymptotic past of a scalar field with {\it constant mass}. This can be 
approximated by the simpler expressions  
\beq 
\label{eq:ini} 
u_k(\taui)= {1\over \sqrt{2\omega_k(\taui)}}, 
\quad   
u_k'(\taui)=-i \sqrt{\omega_k(\taui)\over 2}, 
\eeq 
which correspond to an initial {\it adiabatic} vacuum, with the  
(arbitrary) initial phase set to  zero.  
 
\subsection{Numerical results} 
 We have computed numerically, for various  
sets of parameters, the asymptotic spectrum of fluctuations, defined as  
the limit when $\tau \rightarrow 0$, i.e. at late times, of the  
expression (\ref{eq:power_spectrum}). This is obtained by solving   
numerically the equation 
\beq 
{d^2 u_k\over d\tau^2}+\left\{k^2-{2\over \tau^2} 
-4 \m^2 \qi H\tau\,  \cos^2\left[{{\m}\over H}\ln(-H\tau)+\varphi\right]   
\right\} u_k=0, 
\label{u}
\eeq 
where we have introduced the notation
\beq 
\qi \equiv \frac{g^2 \Phii^2}{4 \m^2}, \label{eq:qi_def} 
\eeq 
and with the  initial conditions defined in  
(\ref{eq:ini}). 
 
Note that this equation only depends on two parameters, for example  
$\qi$ and the ratio $\m/H$. If one changes $\m$ while keeping those two  
parameters fixed, the modification of the spectrum  
 simply corresponds to a rescaling of the wavenumber $k$, as one can  
check explicitly by dividing  equation (\ref{u}) by $\m^2$ in order  
to work with the dimensionless variable $\m\tau$.

\begin{figure} 
\begin{center} 
\includegraphics[width=3.5in]{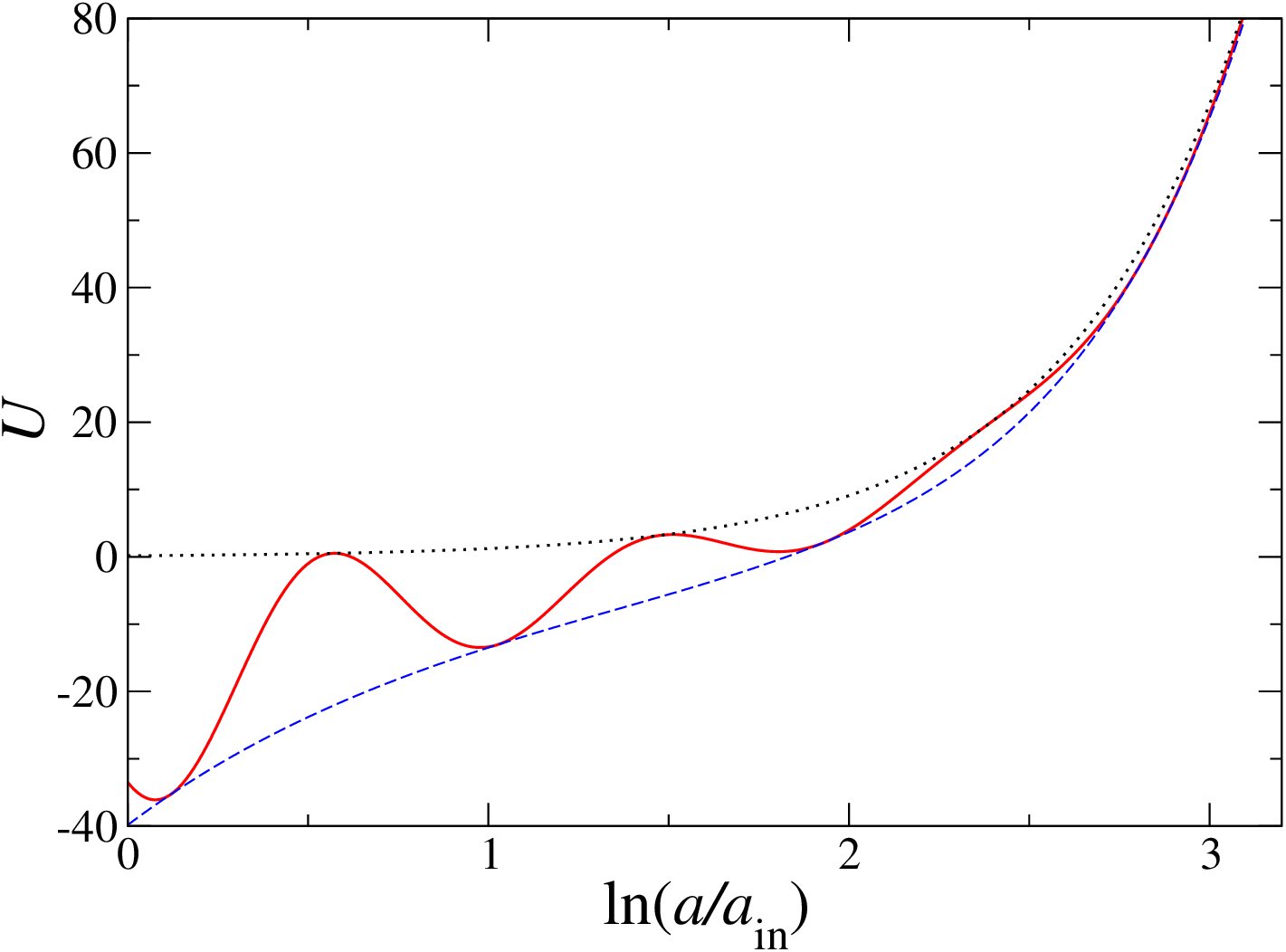} 
\end{center} 
\caption{Effective potential for 
$u_k$ during de Sitter inflation (solid line), as a function of the number of
e-folds $\ln(a/a_{\rm in})=-\ln(-H\tau)$.  
For comparison, the oscillation-free  potential  
Eq.~(\ref{potential_smooth}), and  
the standard de Sitter inflationary 
potential without coupling, $2/\tau^2$, are also shown (dotted line 
and dashed line, respectively). The parameters are   
$\qi=10$  and $\m /H=2 \sqrt{3}$. 
}  
\label{fig:potential_pap} 
\end{figure}

As an example, we have plotted (solid line)  
in Fig.~\ref{fig:potential_pap} the effective 
potential for  the parameters $\qi=10$  
and $\m/H=2\sqrt{3}$. 
For comparison,  
we have also plotted  
the pure de Sitter effective potential (dotted line), i.e.,  
 without coupling to $\phi$, and the effective  potential  
{\it without oscillations} (dashed line), i.e. 
\beq 
\label{potential_smooth} 
\V(\tau)= {2\over \tau^2} 
+g^2\Phii^2 H\tau\, . 
\eeq 
 
\begin{figure} 
\begin{center} 
\includegraphics[width=3.5in]{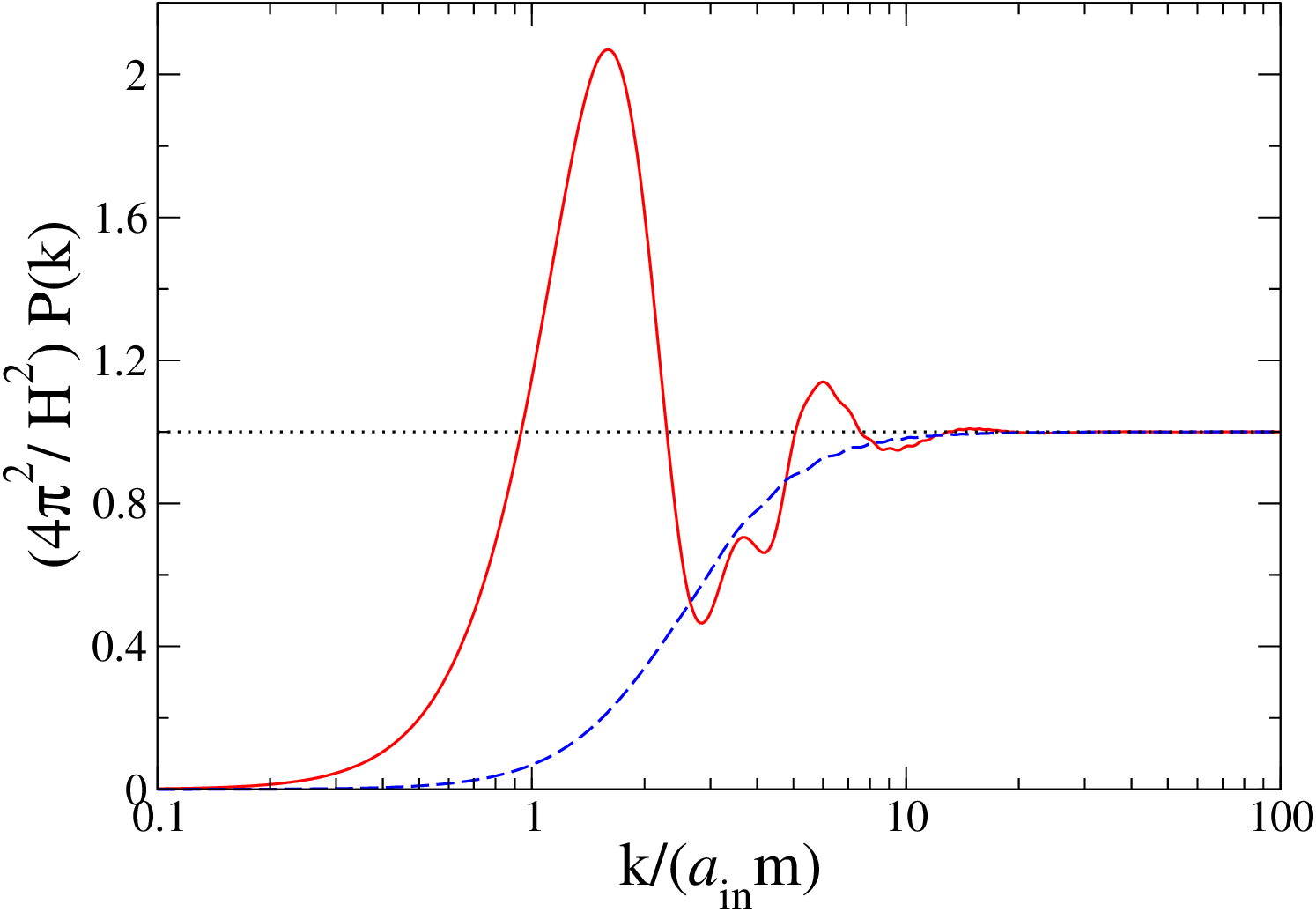} 
\end{center} 
\caption{Power spectrum of the field $\chi$, where the evolution 
of $u_k = a \chi_k $   
is governed by the three potentials of Fig.~1, respectively.}  
\label{fig:powerspectrum} 
\end{figure}

The corresponding power spectra are  plotted in Fig.~\ref{fig:powerspectrum}.
For the effective potential without oscillations,  
one observes a smooth transition between the spectrum of  
a very massive scalar field, which is suppressed, and the standard  
scale invariant spectrum of a massless scalar field.  Indeed, for
small $k$, i.e., large scales,  the modes leave  the Hubble 
radius  when   
the effective mass  is large, thus suppressing the power spectrum  
on large scales. 
However, as  the effective mass  decays with time, the amplification 
 due to inflation  becomes efficient. 
Since the $g^2 \phi^2$ term drops down very rapidly, the transition 
between the strongly suppressed spectrum and the flat spectrum produced  
at earlier stage by inflation occurs very rapidly, within a single  
e-fold of inflation. 
 
By contrast, the spectrum with an oscillating $\phi$ shows a much  
 more complicated transition between a suppressed spectrum and the usual  
massless spectrum. One finds a strongly {\it modulated} spectrum with  
 peaks and troughs. The amplitude  
of some peaks can be even  
larger than the amplitude for a massless scalar field. 
As we will discuss soon, this complicated spectrum is due to the combination 
of two opposite effects: the suppression due to a significant effective  
mass, the amplification due to parametric resonance.  
 
\begin{figure} 
\begin{center} 
\includegraphics[width=4in]{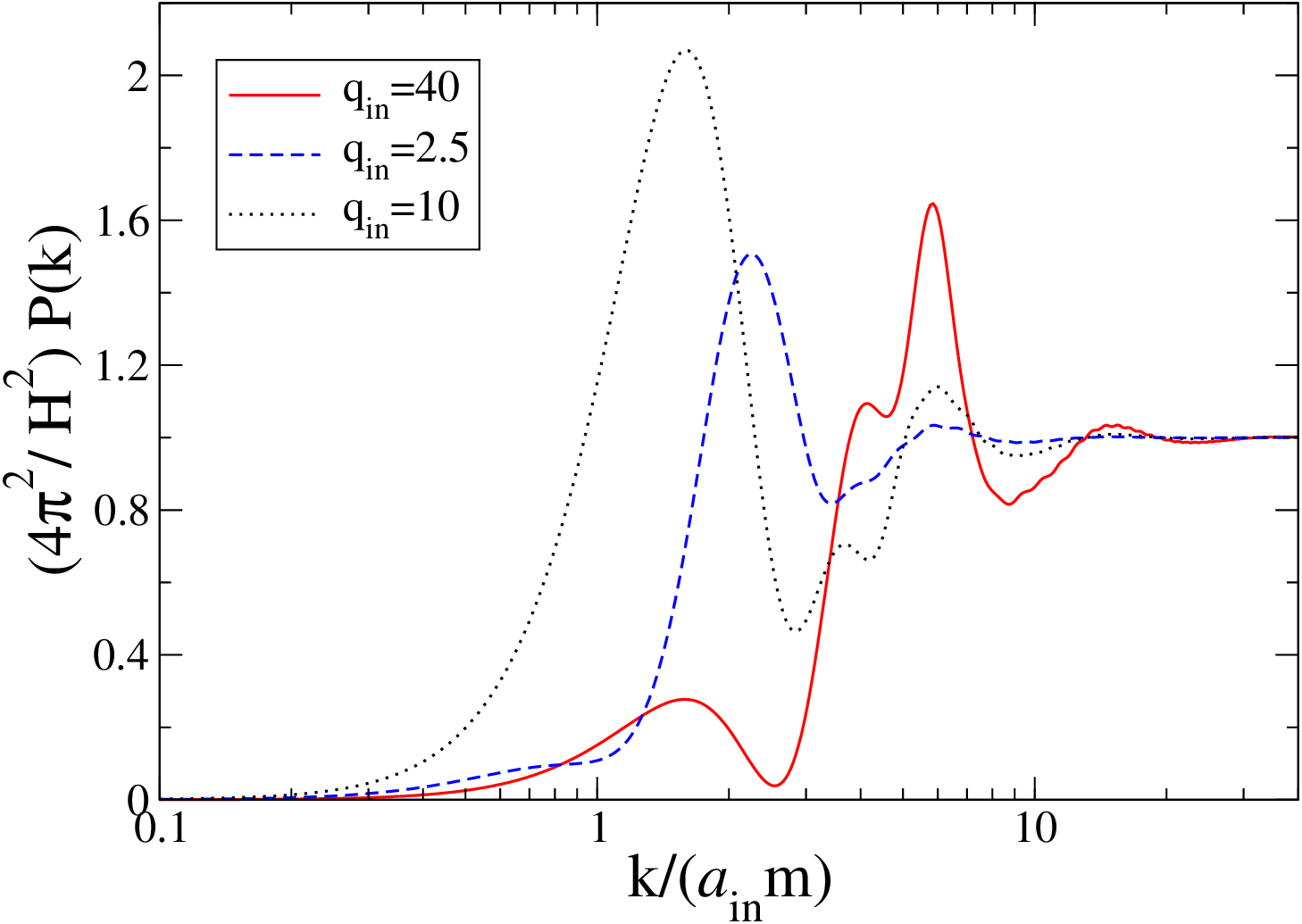} 
\end{center} 
\caption{Spectra for various values of $\qi=g^2 \Phii^2 /(4 \m^2)$, with  
${\m}/H=2 \sqrt{3}$. 
}  
\label{eq:varyingq} 
\end{figure}

\begin{figure} 
\begin{center} 
\includegraphics[width=4in]{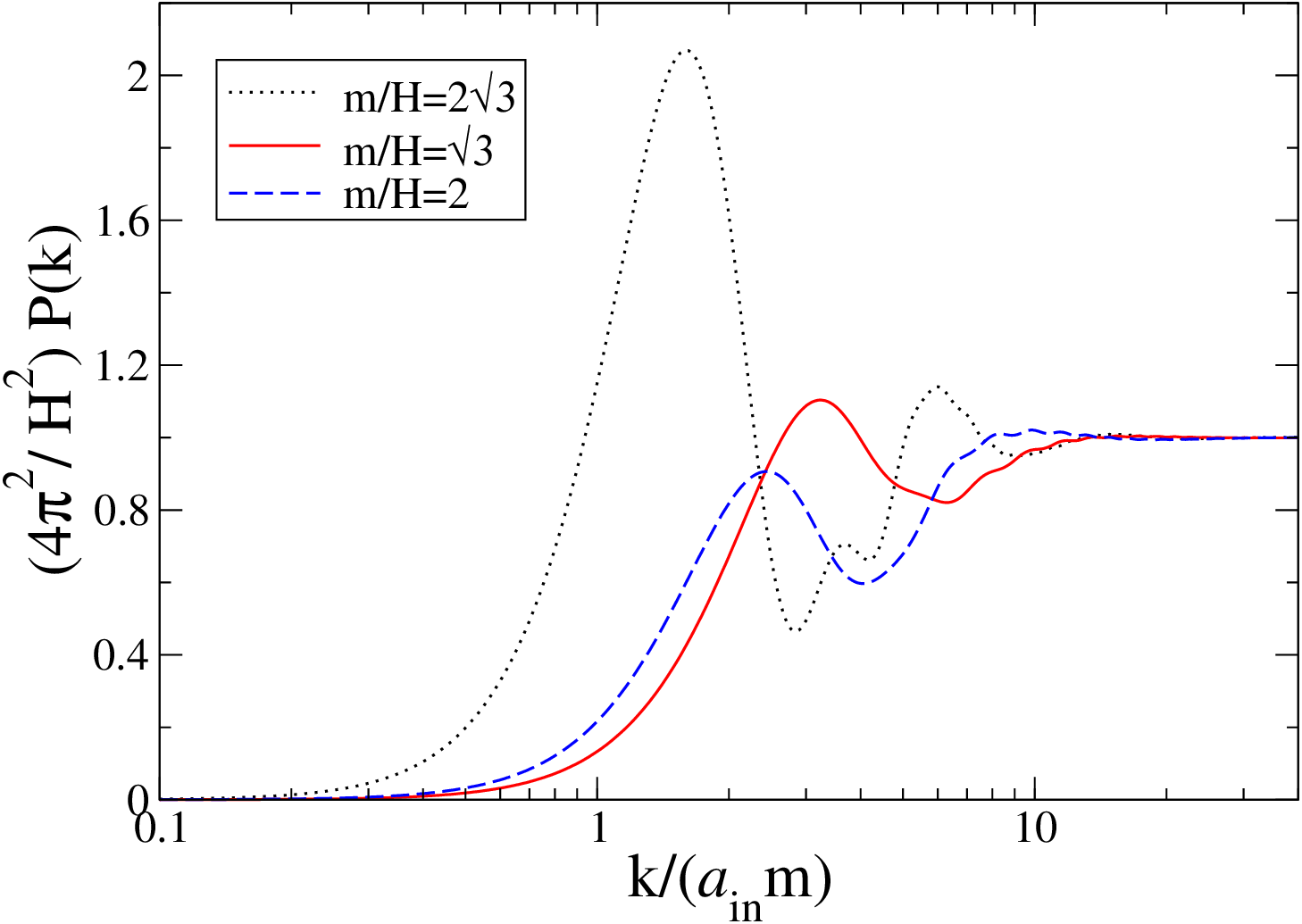} 
\end{center} 
\caption{Spectra for various values of  
${\m}/H$, with $\qi=g^2 \Phii^2 /(4 \m^2)=10$. 
}  
\label{eq:varyingH} 
\end{figure}

In Fig.~\ref{eq:varyingq}, we also show  
the power  spectra for  different values of the 
parameter $\qi$, whereas Fig.~\ref{eq:varyingH} contains the 
power spectra for different values  
of $\m/H$.  
 
\subsection{Discussion} 
In order to understand the spectra that we have computed, it is instructive 
to attempt an analysis similar to the preheating scenario. A key  
observation is that particle production occurs when the evolution  
is non-adiabatic, i.e., for  
\beq 
\label{non-adiab} 
\left| {d\omega_k\over d\tau}\right| \gsim \omega_k^2.  
\eeq 
In the present case, there are two possible sources for the  
non-adiabati\-city. 
At early times, this occurs when $\phi$ is near the origin and corresponds 
to a local maximum of the effective potential $\V(\tau)$. At later  
times, for scales larger than the Hubble radius,  
 adiabaticity is violated because of the de Sitter curvature. The  
first source of non-adiabaticity is the mechanism for particle production in 
the preheating scenario, whereas the second source is the mechanism  
responsible for the amplification of the vacuum fluctuations during   
inflation. In our case, the two effects above are combined and their   
relative influence on the fluctuations will be scale-dependent. 
 
One can estimate which modes will be affected by the amplification. 
Substituting the effective frequency in Eq.~(\ref{non-adiab}) 
and  
using $\phi' = a  \dot \phi 
\simeq a  \m \Phii a^{-3/2}$ (assuming that $\m \gg H$), 
one finds that adiabaticity is violated  when 
\beq 
k^2 \lsim  
\left( a^{3/2} g^2 \phi \m \Phii+ \frac{2}{\tau^3} \right)^{2/3}  
- g^2 \phi^2 a^2  
+\frac{2}{\tau^2}.  
\eeq 
The amplitude of the oscillations in the potential decreases exponentially 
with time 
and after a few oscillations the effect of inflation, which depends 
on the term $2/ \tau^2$, dominates over 
the coupling to $\phi$. 
The moment when this happens  can be estimated by requiring 
\beq 
a^{3/2} g^2 \phi \m \Phii< - \frac{2}{\tau^3} . \label{infla_over_phi} 
\eeq 
[For $\m>H$, Eq.~(\ref{infla_over_phi})  
also implies that $g^2 \phi^2 a^2 < 2/ \tau^2$.]

The potential 
$\V(\tau)$, shown in Fig.~\ref{fig:potential_pap},   
 possesses, at early times,  local maxima  
corresponding to the zeros of  $\phi$, i.e. for 
$t_j = \frac{\pi}{\m} \left(j - \frac{1}{2} \right)$, or 
$\tau_j =-H^{-1} \exp(-H t_j)$, 
for $j=1, 2, \ldots$.  
Using  Eq.~(\ref{infla_over_phi}) 
one sees that beyond the zero of $\phi$ labeled by   $j_{\rm max}$ 
defined as the integer part of  
\beq 
J= \frac{1}{2} + \frac{1}{3 \pi} \frac{\m}{H }  
\ln   
\left[ 2 \qi \left( \frac{\m}{H} \right)^3 \right], 
\eeq 
the de Sitter term always dominates the term due  the coupling to $\phi$.  
During the first few $j$ oscillations, with $j< j_{\rm max}$,  
the main effect is the oscillating mass. Ignoring the de Sitter term,  
 one finds that adiabaticity is violated  when 
\beq 
k^2 \lsim a (g^2 \phi \m \Phii)^{2/3} - g^2 \phi^2 a^2 . \label{eq:adiab1} 
\eeq 
This condition can be  satisfied for small $k$ when the right hand side  
is positive, which implies that $\phi$ must be sufficiently close  
to the origin so that   
\beq 
\label{cond_preheating} 
\phi\lsim  a^{-3/4} \sqrt{\m \Phii/g}.  
\eeq  
We have ignored here the de Sitter term. There is however a subtlety:  
when $\phi$ is  
very close to zero, then  the de Sitter term necessarily dominates in the  
effective mass. When the condition (\ref{cond_preheating}) is saturated,  
this is the case if  
\beq 
a\gsim \left({ \m\over H}\right)^{4/3} \qi^{1/3}. 
\eeq 
The de Sitter term brings a negative  
contribution to the effective square mass. Therefore, its  
effect when $\phi$ is close to zero is indeed 
to amplify the preheating-type effect.

From Eq.~(\ref{eq:adiab1}), during the $j$th oscillation, 
the maximal range of momenta where we have particle production 
due to the coupling to $\phi$ 
is then $k\leq k^{(j)}_*$, where 
\bea 
k^{(j)}_* &\equiv & a^{1/4}(\tau_{j}) \sqrt{g \m \Phii} \\ 
&=& \sqrt{2} \m \left( a(\tau_{j}) \qi \right)^{1/4} 
, \quad  
{\rm for  } \ \  j < j_{\rm max}. \label{eq:k_star} 
\eea 
In conclusion, modes with $k>k_{\rm max}$, 
\beq 
k_{\rm max} \equiv \sqrt{2} \m 
\left( a(\tau_{j_{\rm max}}) \qi \right)^{1/4}, \label{eq:max_mode} 
\eeq 
only feel the effect of inflation. 
We can use this equation  to  
estimate the largest wavenumber for which  
we get a  modulation in the spectrum and we find  
a very good agreement with our numerical results. 
 
For modes with small $k$ the combined effects of, first,  
parametric resonance (possibly amplified by the de Sitter effect) and,  
later, de Sitter amplification can lead either to an enhancement or  
to a suppression, 
 as illustrated by the strong modulation of the spectra on large scales.  
 The  effects are subtle. For example, increasing the  
value of $\qi$  tends to make the parametric resonance
 more efficient  but 
at the same time, a large $\qi$ implies that the effective mass  
will be on average larger and can thus lead to a suppression of the power  
spectrum 
generated by inflation.  

\begin{figure} 
\begin{center} 
\includegraphics[width=4in]{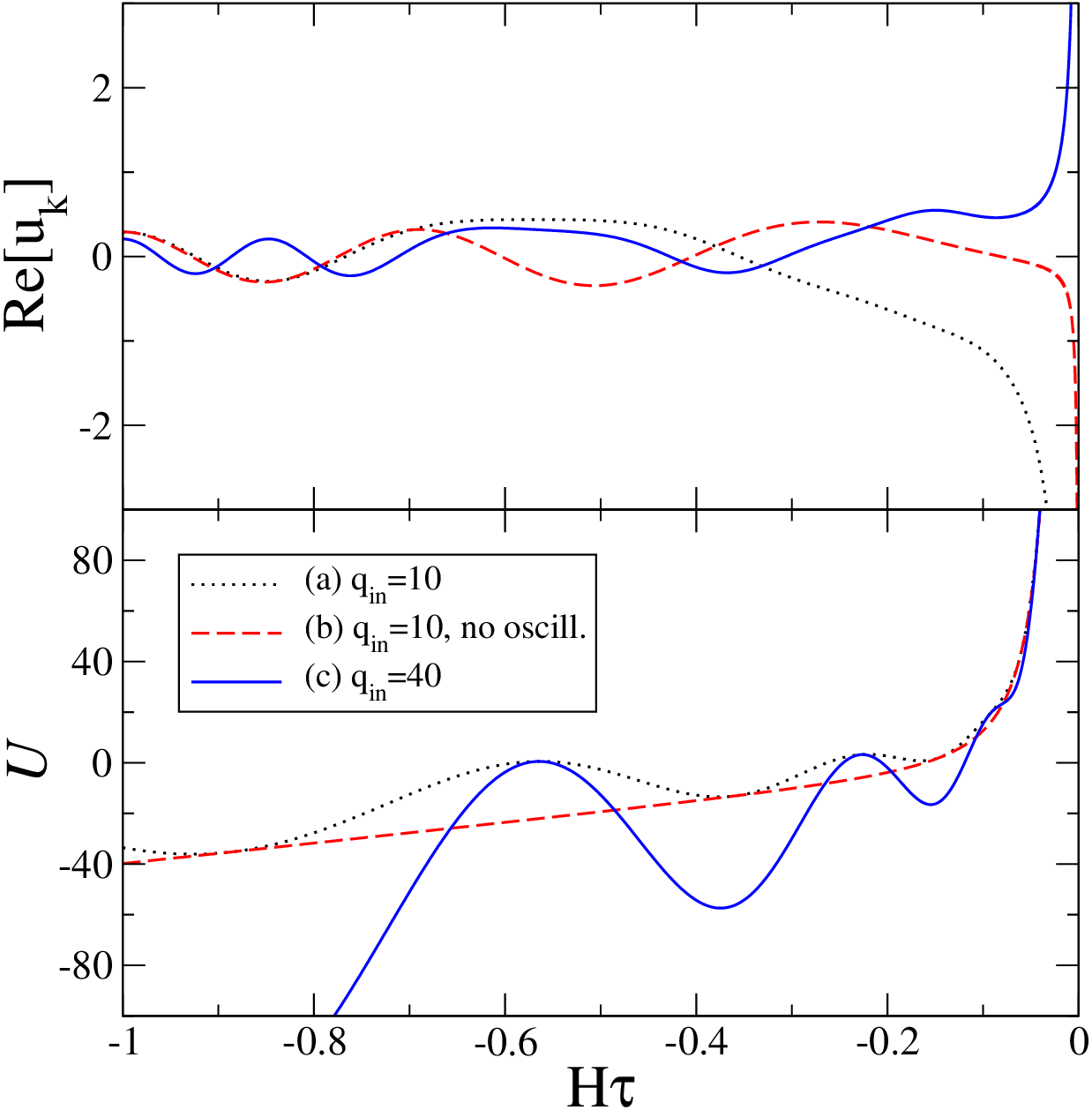} 
\end{center} 
\caption{Upper panel: evolution of the mode $k=1$ in three cases;  
(a): $\qi=10$; (b): $\qi=10$,  
 with the oscillation-free potential of  Eq.~(\ref{potential_smooth}),  
and (c): $\qi=40$. 
Lower panel:  
the effective potential for the three cases.}  
\label{fig:real_pap} 
\end{figure}

To illustrate how the various effects can combine, we have plotted  
in Fig.~\ref{fig:real_pap} 
 the time evolution of the mode $k=1$ in three different cases: 
(a): $\qi=10$; (b): $\qi=10$,  
but with the oscillation-free potential
 given in Eq.~(\ref{potential_smooth}),  
and (c): $\qi=40$. 
In parallel, we have  
plotted the effective potential $\V(\tau)$ for the three cases.  
From the spectra plotted in Fig.~2 and 
Fig.~3 we know that the case (a)  corresponds 
to a strong enhancement while the case (c) is strongly suppressed and  
(b) even more. The time evolution is instructive: the case (b) oscillates  
regularly and is strongly amplified at late times by the de Sitter effect, 
although it has not been amplified by the parametric resonance before. 
In the cases (a) and (c), the effect of the oscillating mass is clearly  
visible. When the effective mass approaches zero, near $H\tau=-0.6$, the  
oscillation of the modes is suspended and their amplitude slightly increases. 
One also sees why the case (c) corresponds to a strong suppression. This  
is because the effective mass, still important near $H\tau=-0.23$,  
exerts a restoring force which reduces  the amplitude before the final  
increase due to the de Sitter amplification.

\section{Power-law inflation} \label{sec:5} 
So far, we have assumed that the  inflationary phase was described by  
a pure de Sitter phase, characterized by a constant $H$. This implies that  
the scalar field has always been oscillating in the past and we were obliged 
to start the computation, somewhat artificially, at a local extremum  of  
$\phi$.  
More realistically,  the Hubble parameter is expected to decrease 
 during inflation. This significantly  changes the past of the scenario  
since  the  massive scalar field $\phi$  
is frozen  
during the  early stage of inflation when $H\gg \mo$.  
Later, when the Hubble parameter  
becomes smaller than its mass, $\phi$ 
 starts to oscillate.

\begin{figure} 
\begin{center} 
\includegraphics[width=4in]{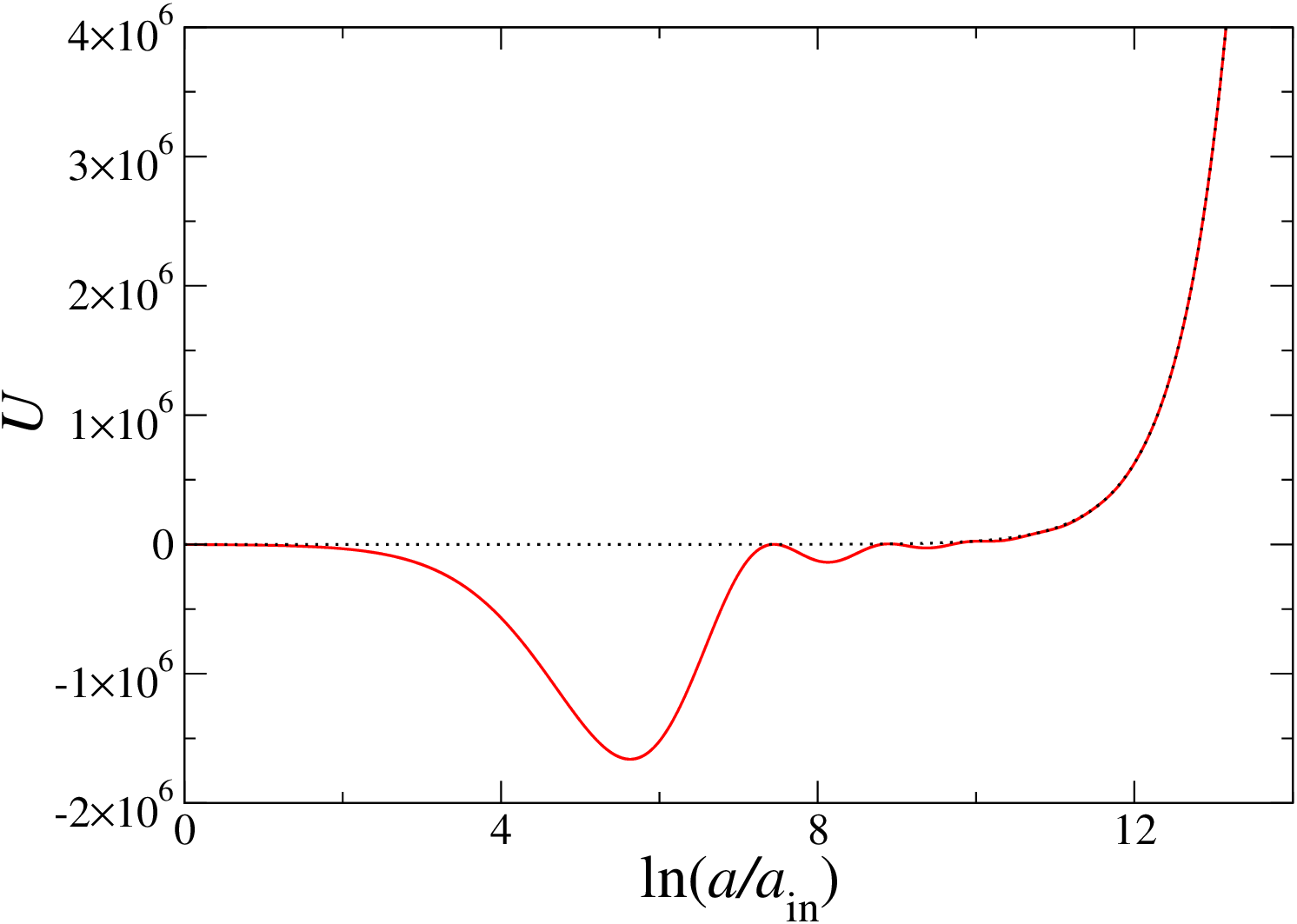} 
\end{center} 
\caption{ 
Potential $U$  for  power law-inflation, with $p=5$, 
where $a \propto t^p$. 
For this example we have used $\qi=6.25\times 10^{5}$.}  
\label{fig:potential_pp_pap} 
\end{figure} 
\begin{figure} 
\begin{center} 
\includegraphics[width=4in]{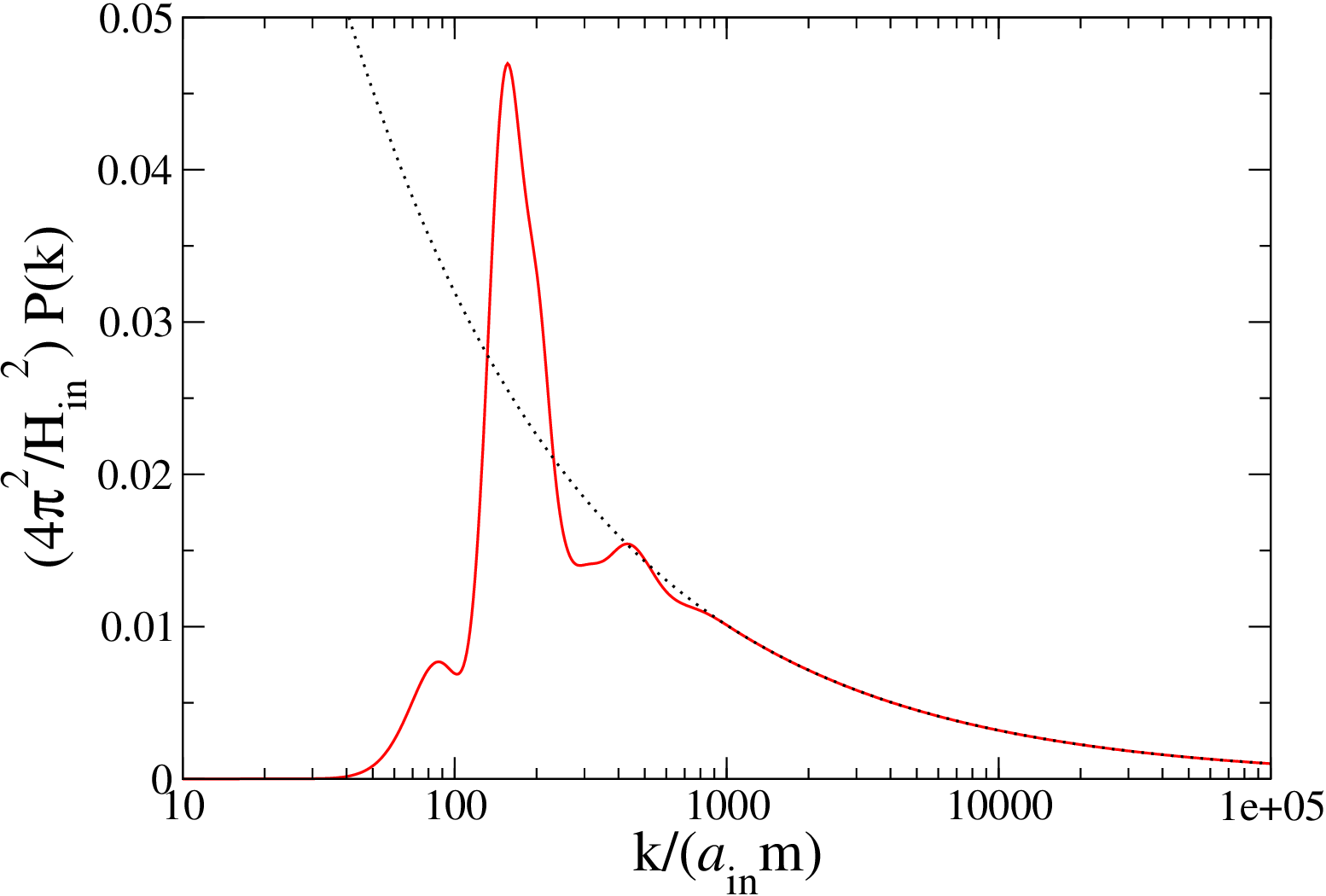} 
\end{center} 
\caption{Power spectrum of the field $\chi$ corresponding to the 
effective potential $U$ of 
 Fig.~\ref{fig:potential_pp_pap}. }  
\label{fig:powerlaw_pap} 
\end{figure} 
 
In order to implement  this more realistic scenario, one can assume that the  
universe undergoes a power-law inflation \cite{LM}, so that the scale factor  
evolves like 
\beq 
a(t)\propto t^p, \qquad p\gg 1. 
\eeq 
Then the equation of motion for $\phi$ reads 
\beq 
\ddot\phi+ {3p\over t} \dot\phi+ \mo^2\phi=0, 
\eeq 
and the 
 solution of this equation that does not diverge when $t\rightarrow 0$ is 
given by  
\beq 
\phi=\phi_0 \Gamma(\nu +1)  
\left({\mo\, t\over 2}\right)^{-\nu}J_\nu(\mo\,  t), \quad 
\nu= (3p-1)/2, 
\eeq 
where $\phi_0$ is the asymptotic value of $\phi$ when  
$t\rightarrow 0$ and $J_\nu$ 
is the Bessel function of order $\nu$.  
As long as $\mo\,  t\ll 1$, $\phi$ remains frozen 
at the value $\phi_0$. When $t$ is of the order of $\mo^{-1}$, the  
scalar field starts to evolve and subsequently oscillates while its amplitude 
decreases like $a^{-3/2}$.  
 
We have plotted in Fig.~\ref{fig:potential_pp_pap} the effective potential 
corresponding to this scenario  and one sees indeed that in the  
asymptotic past $\V(\tau)$ goes to zero, in contrast with the de Sitter  
case of the previous section. 
We have also computed the power spectrum associated with  
this  effective potential in Fig.~\ref{fig:powerlaw_pap}. 
 We observe the same features as in the de Sitter case: a transition 
from a suppressed spectrum to a standard massless spectrum, which is not
scale-invariant here since we are in power-law inflation (the massless 
spectrum is shown as a dotted line), with a strong modulation.

In order to relate the present scenario with that of the previous  
section, it is useful to evaluate the amplitude of the scalar field  
$\phi$ deep in the oscillatory regime where (\ref{sol_phi}) is a  
good approximation as a function of its initially frozen amplitude  
$\phi_0$.  
Using the asymptotic behavior  
$J_\nu(z)\simeq \sqrt{2/(\pi z)} \cos(z-\pi(2\nu+1)/4)$ for $z\gg 1$, one  
can evaluate the amplitude of the scalar field at time $t\gg \mo^{-1}$.  
One finds  
\beq 
\Phi(t)\sim \phi_0 2^{-\nu}\Gamma(\nu+1)\left({\mo p\over H}\right)^{-3p/2}, 
\eeq 
where we have used $t=p/H$.  
This  gives roughly 
\beq 
\Phi(t)\sim \phi_0 \left({3H\over 4 \mo e }\right)^{3p/2}. \label{eq:damping} 
\eeq 
Since $3H/(2\mo)<1$, one sees that the ratio  $\Phi/\phi_0$  
 becomes  extremely small for  
large values of $p$.

\section{Observational consequences} \label{sec:6} 
The purpose of this section is to discuss whether the phenomena studied  
in the present work could be of any relevance  from an observational  
point of view, in particular in the CMB
fluctuations. This is reminiscent of the question whether  preheating
could significantly affect super-Hubble perturbations \cite{Preheatingcurvature}.
 
Let us first stress  that,  
since $\chi$ is not the inflaton in our model,  
its spectrum will not be directly relevant   
if the CMB fluctuations observed today are essentially due to the  
inflaton fluctuations. However, this is not necessarily the case as  
demonstrated recently by  the curvaton \cite{lw02,es02,mt01} 
and inhomogeneous reheating scenarios \cite{Dvali:2003em,Kofman:2003nx,Io}, 
in which the primordial perturbations are generated by a light field that is 
not driving inflation.

Could, then,  our field $\chi$ be a curvaton?  
If one starts from the potential  
given in Eq.~(\ref{lagrangian}), it does not seem to be a viable idea. 
Indeed, as mentioned earlier, the background value of $\chi$ has been driven  
to zero by its initially large effective mass and, even after $\phi$ has fallen 
to zero, there is a priori  
no reason for $\chi$ to acquire a non-zero expectation  
value. This is problematic because  
 the energy density fluctuations due to the scalar field  
$\chi$ will be, instead of linear, quadratic in $\delta \chi$ and will not  
have the required Gaussian properties \cite{lw02}. 
 
This objection, however, can be evaded by a slight modification of the  
scalar field potential, which does not change the power spectrum generated 
during inflation.  
Let us consider  a potential of the form 
\beq 
V(\phi,\chi)= 
{\textstyle {1\over 2}} \mo^2\phi^2 +{\textstyle {1 \over 2}} g^2 \phi^2  
\left(\chi-\chi_0\right)^2 
+ {\textstyle {1 \over 2}} m_\chi^2 \chi^2, 
\eeq 
where $m_\chi$ is a mass scale much smaller than the Hubble parameter  
during inflation, as required for a curvaton. The above potential has
different minima for $\chi$, depending on the value of $\phi$. In the  
very early universe, when the value of $\phi$ is big, $\chi$ is 
driven towards
 the  value $\chi_0$ because of the second term in the potential. 
The field  
$\phi$  subsequently oscillates towards zero. However, because of the  
Hubble friction, the 
scalar field $\chi$ remains frozen at its fixed value $\chi_0$ for a while.  
Eventually, when $H$ reaches down the value  $m_\chi$,  
$\chi$ starts oscillating about its low-energy minimum  
$\chi=0$, as a standard curvaton. 
If $\chi$ oscillates for long enough before
decaying, it will eventually dominate the energy density of the universe,
because its energy density decreases more slowly that the radiation
produced by the inflaton decay. This implies that the isocurvature $\chi$
fluctuations are converted into curvature perturbations \cite{lw02},
\beq
\zeta = \frac{2}{3} \frac{\delta\chi_*}{\chi_*},
\eeq
where, $\zeta \equiv 
-H \delta \rho /\dot \rho $ is the uniform energy density perturbation,
here written in the spatially flat gauge. 
The star refers to the expectation value of $\chi$ during inflation, and thus
$\chi_*=\chi_0$, and $\delta\chi_*$ corresponds to the fluctuations
generated during inflation.

In this context, one sees that $\chi$ behaves exactly as a curvaton, with the 
only exception that  its quantum fluctuations during inflation are  
affected by the evolution of the scalar field $\phi$. Therefore, if the  
usual conditions for neglecting the inflaton perturbations with respect  
to the curvaton perturbations are satisfied, then  
the CMB fluctuations would contain the special features 
 discussed in the previous  
sections. One could also envisage the possibility that the primordial  
spectrum is the combination of a standard inflaton spectrum and of a  
modulated spectrum of the kind discussed here \cite{lv04a}. 
 
One legitimate question is then whether the peculiar features of our  
modulated spectra could account for the ``anomalies''  that   
seem to be suggested by the WMAP data \cite{Bennet,Spergel}:  
the so-called cosmic variance outliers  
(points which lie outside the one sigma cosmic variance)  
are present either on large scales, in the suppression of the  
lower multipoles, and on smaller scales,  where the power spectrum 
seems to contain features in the form of spikes or waves. Several works 
(see e.g. \cite{Bridle,Peiris,Contaldi,Julien} 
for the lowest  
multipoles and \cite{Martin,Hunt:2004vt} for higher  
multipoles) have suggested various modifications  
of the early universe scenario  to account for this. 
 
\begin{figure} 
\begin{center} 
\includegraphics[width=5in]{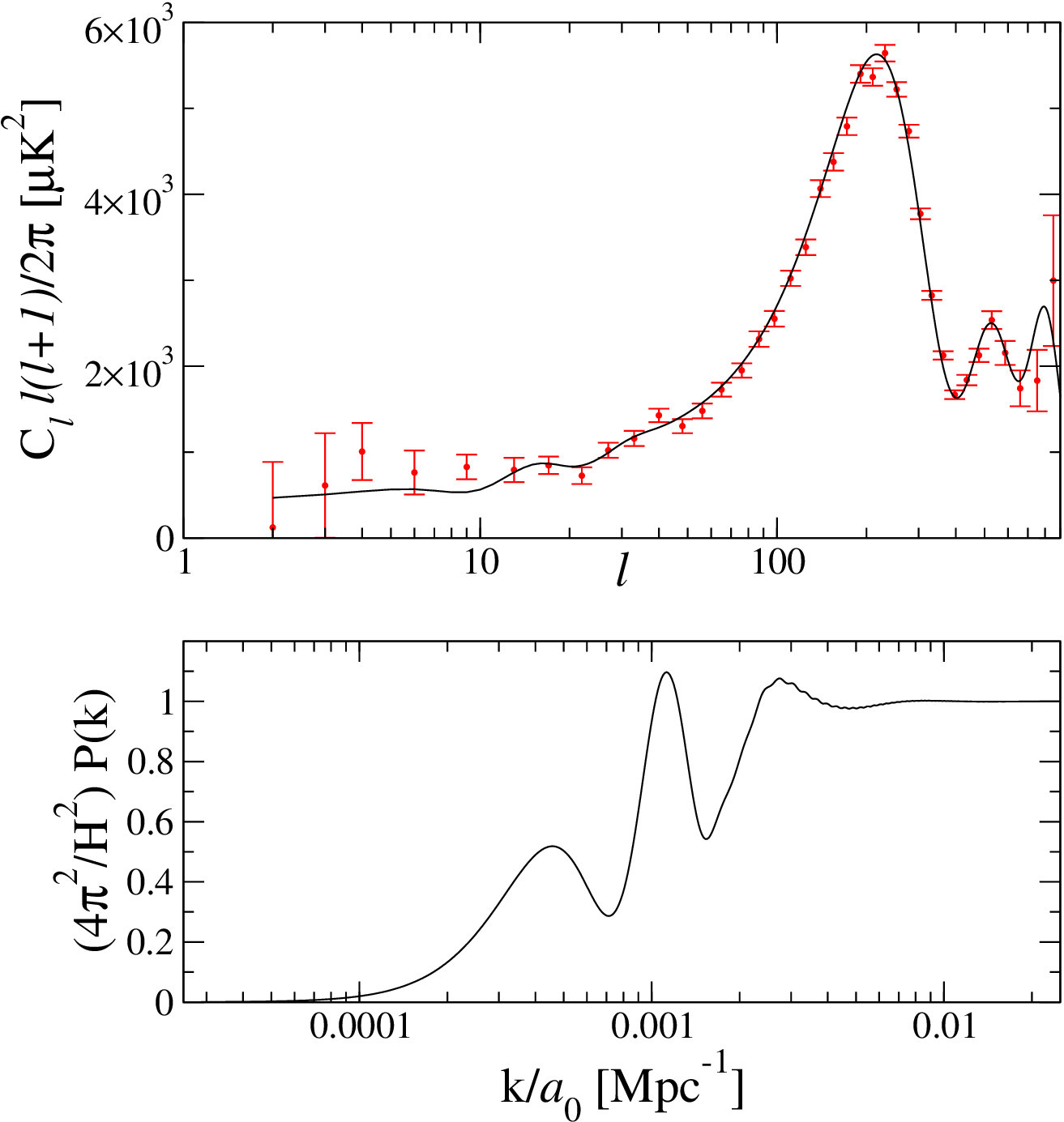} 
\end{center} 
\caption{In the upper panel,  
an example of spectrum of CMB anisotropies generated by  
a primordial curvaton power spectrum with modulated fluctuations, 
computed by modifying CAMB \cite{Lewis}.  
We use the minimal set of $\Lambda$CDM model parameters from the best fit  
to the WMAP data alone as provided by \cite{Spergel}. 
No running of the spectral index or tensor  
contribution have been included.  
The latter would rise the CMB spectrum on large scales helping  
to fit the data. 
The spectrum is compared to the first year WMAP data. In the lower panel the  
corresponding power spectrum of $\chi$, generated by using the parameters 
$\qi=40$ and $\m/H=2 \sqrt{2}$.} \label{fig:Cl} 
\end{figure} 
 
It is not the purpose of the present work  to investigate systematically  
this question. We will just limit ourselves to a few remarks.   
The power spectrum of $\chi$ as derived in Sec.~\ref{sec:4} 
has two characteristics: a strong suppression  
on very large scales and a series of few oscillations on slightly  
smaller scales. On smaller length-scales, for $k \gsim k_{\rm max}$, it 
 is  featureless and close to scale invariant. 
In \cite{Bridle} it was noticed  that the WMAP data  
are consistent with a primordial power spectrum which is featureless for  
the modes $k/a_0 \gsim 0.005 \ {\rm Mpc}^{-1}$. Thus the modulations  
should affect modes with larger length-scales. 

As an illustration, we have constructed such an example
 of  modulated power spectrum, shown 
in the lower panel of Fig.~\ref{fig:Cl},  
with the parameters $\qi=40$ and $\m/H=2 \sqrt{2}$. In the upper  
panel, we have plotted the corresponding spectrum for CMB anisotropies,  
computed by modifying CAMB \cite{Lewis}.  
The large scale plateau of the CMB anisotropies 
is suppressed for $\ell \lsim 30$,  
corresponding to the suppression of the power spectrum  
of $\chi$, on very large scales, $k/(a_0 {\rm Mpc}^{-1}) 
 \lsim  10^{-3}$. 
Furthermore, the  
oscillations in the $\chi$ spectrum for the intermediate scales
translate into oscillations 
in the CMB spectrum for $10 \lsim \ell \lsim 40$.

Note however that in order for these features to be observable, the  
modulations due to the oscillating mass must affect, during inflation,  
precisely the scales that have reentered the Hubble radius recently, i.e.  
the scales corresponding to the largest observable scales. A similar  
problem hampers the models trying to  explain particular features  
with a modification of the early universe dynamics.

\section{Conclusions}
In the present work, we have investigated the amplification of 
the quantum fluctuations of a test scalar field $\chi$ during inflation, when 
this scalar field has an oscillating effective mass, resulting from 
its coupling to a massive  scalar field $\phi$ 
that oscillates during inflation. 

We have computed numerically the final power spectrum for the fluctuations, 
which in the context of de Sitter inflation, depends only on two parameters:
the ratio $m_\phi/H$ and the (initial) coupling parameter 
$\qi=g^2\Phii^2/(4m^2)$. The power spectrum is characterized by a transition
from a suppressed spectrum on large length-scales to a standard massless 
spectrum on smaller length-scales. On the intermediate scales, the 
spectrum can be affected by strong modulations.  

Our scenario 
shares with the preheating scenario the essential ingredient of an 
  effective oscillating mass. However, in our case, the universe is 
accelerating whereas, in the preheating scenario, taking place after inflation,
the universe is decelerating. As a consequence, instead of particle 
production in the usual sense,  the  fluctuations, in our case, 
 are amplified on super-Hubble 
scales. Moreover, in addition to the parametric resonance due the 
oscillating mass, our scenario is also characterized by an amplification due 
to the acceleration of the background cosmology. 

We have also discussed a possible observational signature of our scenario 
when the scalar field $\chi$ is assumed to be a curvaton field. We have 
shown that the corresponding 
CMB multipole spectrum is  characterized by a suppression 
on large angular scales and a modulation on slighter smaller scales, with
respect to the standard power spectrum.

\vskip 1cm 
{\bf Acknowledgements} 
We wish to thank Justin Khoury for 
stimulating discussions and instructive comments.  F.V. acknowledges  
financial support from the Swiss National Science Foundation.

\end{document}